\documentclass[12pt]{amsart}
\usepackage{graphicx,amsmath,amssymb}

\newcommand{\C}{\mathbb C}
\newcommand{\R}{\mathbb R}

\newcommand{\N}{\mathbb N}
\renewcommand{\d}{\prime} 
\newcommand{\dd}{{\prime \prime}}

\renewcommand{\Re}{{\rm Re}\,}
\renewcommand{\Im}{{\rm Im}\,}
\newtheorem{theorem}{Theorem}
\newtheorem{lemma}[theorem]{Lemma}
 
\newtheorem{corollary}[theorem]{Corollary}

\newtheorem*{remark}{Remark}

\begin{document}
\title[]
{On the shape of spectra for non-self-adjoint\\
 periodic Schr\"odinger operators}
\author[]
{Kwang C. Shin}
\address{Department of Mathematics, University of Missouri, Columbia, MO 65211}
\date{April 5, 2004}

\begin{abstract}
The spectra of the Schr\"odinger operators with periodic potentials are studied. When the potential is real and periodic, the spectrum consists of at most countably many line segments (energy bands) on the real line, while when the potential is complex and periodic,  the spectrum consists of at most countably many analytic arcs in the complex plane. 

In some recent papers, such operators with complex $\mathcal{PT}$-symmetric periodic potentials are studied. 
In particular, the authors argued that some energy bands would appear and disappear under perturbations. 
Here, we show that  appearance and disappearance of such energy bands imply existence of nonreal spectra. This is a consequence of a more general result, describing the local shape of the spectrum. 
\end{abstract}

\maketitle

\baselineskip = 18pt
 In recent papers \cite{Ahmed,Bender,Jose,Jose1,Jones}, appearance and disappearance of real energy bands for some complex $\mathcal{PT}$-symmetric periodic potentials under perturbations have been reported. In this paper, we show that appearance and disappearance of such real energy bands imply existence of nonreal band spectra.

We begin by introducing some facts on Floquet theory and the associated Hill operators $H$. Consider the Schr\"odinger equation
\begin{equation}\label{main_eq}
-\psi_{xx}(E,\,x)+V(x)\psi(E,\,x)=E\psi(E,\,x),\quad x\in\R,
\end{equation}
where $E\in\C$ and $V\in L^1_{loc}(\R)$ is a complex-valued  periodic function of period $\omega>0$. The {\it Hill operator} $H$ in $L^2(\R)$ associated with
(\ref{main_eq}) is defined by
\begin{align}
(Hf)(x)&=-f_{xx}(x)+V(x)f(x),\nonumber\\
f\in dom(H)&=\left\{g\in L^2(\R)\,:\, g,\, g_x\in AC_{loc};\, (-g_{xx}+Vg)\in L^2(\R)\right\}.\nonumber
\end{align}
Then $H$ is a densely defined closed operator in $L^2(\R)$ (see, e.g.,
\cite[Chap. 5]{MSPE}).

For each $E\in\C$, there exists a fundamental system of solutions $\phi_1(E,\,x),\, \phi_2(E,\,x)$ of (\ref{main_eq}) such that
\begin{eqnarray}
\phi_1(E,\,0)=1,&& \phi_{1\,x}(E,\,0)=0,\nonumber\\
\phi_2(E,\,0)=0,&& \phi_{2\, x}(E,\,0)=1.\nonumber
\end{eqnarray}
The {\it monodromy matrix} $M(E)$ of \eqref{main_eq} is defined by
\begin{equation}\nonumber
M(E)=\left(
\begin{matrix}
\phi_1(E,\omega)& \phi_2(E,\omega)\\
\phi_{1\, x}(E,\omega)& \phi_{2\, x}(E,\omega)
\end{matrix}
\right).
\end{equation}
Then $\rho_1(E)\rho_2(E)=1$, where $\rho_1(E)$ and $\rho_2(E)$ are the eigenvalues (the {\it Floquet multipliers}) of $M(E)$ (see, e.\ g., \cite[eq.\ (1.2.4)]{MSPE}). The {\it Floquet discriminant} $\Delta(E)$ is defined by half of the trace of $M(E)$, that is, by
\begin{equation}\label{trace_eq}
\Delta(E)=\frac{1}{2}\left(\phi_1(E,\,\omega)+\phi_{2\, x}(E,\,\omega)\right)=\frac{1}{2}\left(\rho(E)+\frac{1}{\rho(E)}\right),
\end{equation}
where $\rho(E)\in\C\setminus\{0\}$ is a Floquet multiplier. 
Then $\Delta(E)$ is an entire function of order $\frac{1}{2}$ (see \cite[Chap. 21]{ECT}).

The spectrum $\sigma(H)$ of $H$ is purely continuous and it consists of at most countably infinitely many analytic arcs whose end points $E$ satisfy either $\Delta(E)^2=1$ or $\Delta^\d(E)=0$ \cite{FSRB}, where $\d$ denote $\frac{d}{dE}$. Of course, when $V(x)$ is real-valued, these arcs lie on the real axis. So the spectrum consists of bands on the real axis. 

We will make use of the following lemmas in investigating the shape of the spectrum $\sigma(H)$ when the potential $V(x)$ is complex-valued.
\begin{lemma}\label{lemma_equiv}
The following four assertions are equivalent.
\begin{enumerate}
\item [(i)] $E\in\sigma(H)$.
\item [(ii)]  $|\rho(E)|=1$.
\item [(iii)] $\Delta(E)$ is real and $-1\leq \Delta(E)\leq 1$.
\item [(iv)] Equation (\ref{main_eq}) with $E\in\C$ has a non-constant bounded solution on $\R$. 
\end{enumerate}
\end{lemma}
\begin{proof}
See, for example, \cite[Chaps. 1 and 5]{MSPE} for a proof.
\end{proof}
\begin{lemma}
$$\sigma(H)\subset \{E\in\C\, :\, \Re(E)\geq M_1,\,M_2\leq \Im(E)\leq M_3\},$$
where $M_1=\inf_{x\in\R} \Re(V(x))$,  $M_2=\inf_{x\in\R}\Im(V(x))$ and $M_3=\sup_{x\in\R}\Im(V(x))$.
\end{lemma}
\begin{proof}
Theorem 2.4.2 in \cite{MSPE} and Lemma \ref{lemma_equiv} above imply that $E\in\sigma(H)$ if and only if \eqref{main_eq} has a non-constant solution $\psi(E,\, \cdot)$ with $\psi(E,\,\omega)=e^{it}\psi(E,\,0)$ and $\psi_x(E,\,\omega)=e^{it}\psi_x(E,\,0)$ for some $t\in\R$. Then we multiply \eqref{main_eq} by $\overline{\psi}$, integrate over $[0,\,\omega]$ and use integration by part to get
$$\int_0^{\omega}|\psi_x(E,\,x)|^2\, dx+\int_0^{\omega}V(x)|\psi(E,\,x)|^2\,dx=E\int_0^{\omega}|\psi(E,\,x)|^2\,dx.$$
Then the lemma is an easy consequence of this equation. 
\end{proof}

We next describe the local shape of the  spectrum $\sigma(H)$. 
\begin{theorem}\label{main}
Let $V\in L^1_{loc}(\R)$ be a complex-valued periodic function on $\R$ with period $\omega>0$. 

Suppose  that $\Delta^\d(E_0)\not=0$.\\ 
\noindent (i) If $-1<\Delta(E_0)<1$, then the spectrum near $E_0$ is an analytic arc.

\noindent (ii) If $\Delta(E_0)=\pm 1$, then the spectrum near $E_0$ consists of an analytic arc, one of  whose end points is  $E_0$.

Suppose that  $\Delta^\d(E_0)=\Delta^\dd(E_0)=\ldots=\Delta^{(k-1)}(E_0)=0$ and $\Delta^{(k)}(E_0)\not=0$ for some $E_0\in\C$ and integer $k\geq 2$.

\noindent (iii) If $-1<\Delta(E_0)<1$, then
the spectrum $\sigma(H)$ near  $E_0\in\C$ consists of $2k$-analytic arcs that have a common end point $E_0$. Moreover,  adjacent arcs meet at $E_0$ at an angle of $\frac{\pi}{k}$.

\noindent (iv) If $\Delta(E_0)=\pm 1$, then the spectrum $\sigma(H)$ near  $E_0\in\C$ consists of $k$-analytic arcs that have a common end point $E_0$. Moreover,  adjacent arcs meet at $E_0$ at an angle of $\frac{2\pi}{k}$.
\end{theorem}
\begin{proof}
Since $\Delta(E)$ is an entire function (of order $\frac{1}{2}$) of $E\in\C$, we have that
\begin{eqnarray}
\Delta(E)&=&\Delta(E_0)+\sum_{j=k}^{\infty}\frac{\Delta^{(j)}(E_0)}{j!}(E-E_0)^j\nonumber\\
&=&\Delta(E_0)+\frac{\Delta^{(k)}(E_0)}{k!}(E-E_0)^k\left[1+O(|E-E_0|)\right],\label{delta_eq}
\end{eqnarray}
where $k\in\N$ is the positive integer such that  $\Delta^{(k)}(E_0)\not=0$, $\Delta^{(j)}(E_0)=0$ for all $j=1,\,2,\,\cdots, k-1$.

Suppose that $\Delta(E_0)$ is real. Then 
$\Delta(E)$ is real if and only if $\sum_{j=k}^{\infty}\frac{\Delta^{(j)}(E_0)}{j!}(E-E_0)^j$ is real. Clearly,
\begin{eqnarray}
\Im(\Delta(E))&=&\Im\left(\sum_{j=k}^{\infty}\frac{\Delta^{(j)}(E_0)}{j!}(E-E_0)^j\right)\nonumber\\
&\underset{E\to E_0}{=}&\Im\left(\frac{\Delta^{(k)}(E_0)}{k!}(E-E_0)^k\left[1+O(|E-E_0|)\right]\right).\nonumber
\end{eqnarray}
 So if $\Delta(E)$ is real with $|E-E_0|$ small, then 
\begin{equation}\label{arg_eq}
\arg(E-E_0)\underset{E\to E_0}{=}\frac{1}{k}\left[j\pi-\arg\left(\Delta^{(k)}(E_0)\right)\right]+O(|E-E_0|)\quad j=1,2,\ldots,2k.
\end{equation}

\noindent {\it Proof of (i) and (ii)}: 
Suppose  that $\Delta^\d(E_0)\not=0$. Then assertions (i) and (ii) are easy consequences of a fact from analytic function theory, that is, if $\Delta^\d(E_0)\not=0$ then there is a neighborhood $\mathcal{O}$ of $E_0$ in the complex $E$-plane such that $\Delta(E)\big|_{\mathcal{O}}$ has an analytic inverse function.

\noindent {\it Proof of (iii) and (iv)}:
Suppose that  $\Delta^\d(E_0)=\Delta^\dd(E_0)=\ldots=\Delta^{(k-1)}(E_0)=0$ and $\Delta^{(k)}(E_0)\not=0$ for some $E_0\in\C$ and integer $k\geq 2$. Then we can deduce from \eqref{delta_eq} and \eqref{arg_eq} that when $\Delta(E_0)$ is real, $\Delta(E)$ near $E_0$ is real along $2k$-analytic arcs that have a common end point $E_0$. Also, by \eqref{arg_eq}, one sees that adjacent arcs meet at $E_0$ at an angle of $\frac{\pi}{k}$. By Lemma \ref{lemma_equiv}, all these arcs near $E_0$ belong to the spectrum $\sigma(H)$ if $-1<\Delta(E_0)<1$. This completes proof of (iii). However,  if $\Delta(E_0)=\pm 1$ then by Lemma \ref{lemma_equiv}, every other arc belongs to the spectrum $\sigma(H)$, and hence the adjacent arcs that belong to the spectrum meet at $E_0$ at an angle of $\frac{2\pi}{k}$. This completes proof of (iv). 
\end{proof}

\begin{remark}
{\em Batchenko and  Gesztesy \cite{Fritz} studied  quasi-periodic algebro-geometric KdV potentials and described the corresponding spectrum by using $<g(E,\cdot)^{-1}>=\frac{1}{\omega}\int_0^{\omega}\frac{1}{g(E,x)}\,dx$ where $g(E,\cdot)$ is the diagonal Green's function of $H$.
 In the special periodic case this reduces to $<g(E,\cdot)^{-1}>=-\frac{2}{\omega}\ln\left(\Delta(E)+\sqrt{\Delta(E)-1}\right)$ (see \cite[eq.\ (C.19)]{Fritz}).}
\end{remark}
 
Next, we will show the existence of nonreal spectra of Hill operators with $\mathcal{PT}$-symmetric (i.\ e., $\overline{V(-x)}=V(x)$ for all $x\in\R$) periodic potentials $V$ under some conditions. We will use this to interpret some numerical and analytic studies in \cite{Ahmed,Bender,Jose,Jose1,Jones}.

\begin{corollary}\label{pt_sy}
Suppose that $\overline{V(-x)}=V(x)$ for all $x\in\R$, and suppose that $V$ is periodic.
 If $\Delta(E)$ has a local maximum or minimum at some $E_0\in\R$
 with $-1<\Delta(E_0)<1$, then there must be some nonreal complex numbers $E$ in the spectrum $\sigma(H)$.
\end{corollary}
\begin{proof}
We first observe that when the periodic potential $V$ is $\mathcal{PT}$-symmetric, $\psi(E,\,x)$ is a (bounded) solution of \eqref{main_eq} corresponding to $E$ if and only if $\overline{\psi(E,\,-x)}$ is a (bounded) solution of \eqref{main_eq} corresponding to $\overline{E}$. Then from Lemma \ref{lemma_equiv}  one can show that the spectrum $\sigma(H)\subset\C$ is symmetric with respect to the real axis.  Moreover, $\overline{\Delta(\overline{E})}=\Delta(E)$, and hence $\Delta(E)$ is real for all $E\in\R$.

 Next, since $\Delta(E)$ has a local maximum or minimum at $E_0\in\R$, we have $\Delta^\d(E_0)=0$. Then since $\Delta(E)$ is an entire function of order $\frac{1}{2}$, there exists $k\in\N$  and $k\geq 2$ such that  $\Delta^\d(E_0)=\Delta^\dd(E_0)=\ldots=\Delta^{(k-1)}(E_0)=0$ and $\Delta^{(k)}(E_0)\not=0$.
So by Theorem \ref{main}, $2k$-analytic arcs meet at $E_0$. Since adjacent arcs meet at an angle of $\frac{\pi}{k}$, there exist some nonreal numbers $E$ in the spectrum $\sigma(H)$.  
\end{proof}

\begin{figure}[t]
    \begin{center}
    \includegraphics[width=.4\textwidth]{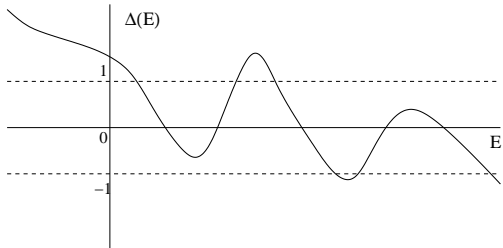}
    \end{center}
 \vspace{-.5cm}
\caption{A sample graph of the Floquet discriminant $\Delta(E)$, $E\in\R$ when $V$ is $\mathcal{PT}$-symmetric.}\label{f:graph1}
\end{figure}

In \cite{Ahmed,Bender,Jose,Jose1,Jones}, Hill operators with $\mathcal{PT}$-symmetric  periodic potentials $V$ are studied. More precisely, the potentials $V(x)=i\sin^{2n+1} x$ are considered  in \cite{Bender}, while $\mathcal{PT}$-symmetric potentials with some delta distributions are studied in \cite{Ahmed,Jose,Jose1,Jones}. The authors in \cite{Ahmed,Bender,Jose,Jose1,Jones} presented graphs of $\Delta(E)$, $E\in\R$ similar to Figure \ref{f:graph1}  and argued that some of the energy bands appear and disappear under  perturbations. They also argued that some of the (anti)periodic band edges are absent due to the  appearance and disappearance of energy bands. However, they failed to point out  the existence of nonreal spectra and nonreal band edges as a result of complex deformations of real intervals of the spectrum under perturbations.

\begin{figure}[t]
    \begin{center}
    \includegraphics[width=.4\textwidth]{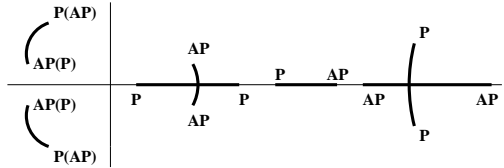}
    \end{center}
 \vspace{-.5cm}
\caption{A {\it possible} spectrum $\sigma(H)$ denoted by dark curves, where anti-periodic and periodic band edges are labeled as AP and P, respectively. The pair of curves on the far left is {\it possible} spectra away from the real axis.}\label{f:graph2}
\end{figure}

The main point of the graph of $\Delta(E)$ in Figure \ref{f:graph1} is that it has local extrema in $(-1,1)$. If this happens, then  we know by Lemma \ref{pt_sy} that nonreal spectra exist. Figure \ref{f:graph2} shows a {\it possible}  spectrum of $H$, corresponding to $\Delta(E)$ in Figure \ref{f:graph1}, where the pair of curves on the far left is  away from the real line. This happens after a real energy band moves off the real axis under  perturbations and  before a new energy band  appears on the real axis.  So the (anti)periodic band edges are not absent in this case, but they become nonreal.

\vskip 5mm

\noindent{\bf Acknowledgments} The author thanks Richard Laugesen for showing him the reference \cite{Bender} that initiated this research and Fritz Gesztesy for pointing out the equation \cite[eq.\ (C.19)]{Fritz} to him.  Also, he thanks Fritz Gesztesy and Richard Laugesen for helpful discussions and suggestions to improve the presentation of this note.

{\sc e-mail contact:}  kcshin@math.missouri.edu
\end{document}